\documentclass[twocolumn,secnumarabic,amssymb, nobibnotes, aps, prd]{revtex4-2}
\usepackage{amssymb}
\usepackage{graphicx}
\usepackage{epsfig}
\usepackage{bigstrut}
\usepackage{amsmath}
\usepackage[mathlines]{lineno}
\usepackage{color}
\usepackage[bookmarksnumbered, pdfstartview=FitH,colorlinks,citecolor=blue,linkcolor=blue,]{hyperref}
\usepackage[perpage,symbol]{footmisc}
\usepackage{subfigure}
\usepackage{indentfirst}
\usepackage{fancyvrb}
\usepackage{multirow}
\usepackage{float}
\setfnsymbol{wiley}

\newcommand{\ra}{\rightarrow}
\newcommand{\jpsi}{J/\psi}
\newcommand{\pio}{\pi^{0}}
\newcommand{\pip}{\pi^{+}}
\newcommand{\pim}{\pi^{-}}
\newcommand{\chisq}{\chi^{2}}
\renewcommand{\thefootnote}{\fnsymbol{footnote}}

\begin{document}

\title{\bf Precision measurement of the matrix elements for $\eta\to\pip\pim\pio$ and $\eta\to\pio\pio\pio$ decays}

\author{
	\begin{center}
M.~Ablikim$^{1}$, M.~N.~Achasov$^{13,b}$, P.~Adlarson$^{73}$, R.~Aliberti$^{34}$, A.~Amoroso$^{72A,72C}$, M.~R.~An$^{38}$, Q.~An$^{69,56}$, Y.~Bai$^{55}$, O.~Bakina$^{35}$, I.~Balossino$^{29A}$
% ...
\\
\vspace{0.2cm}
(BESIII Collaboration)\\
\vspace{0.2cm} {\it
$^{1}$ Institute of High Energy Physics, Beijing 100049, People's Republic of China\\
% ...
$^{13}$ G.I. Budker Institute of Nuclear Physics SB RAS (BINP), Novosibirsk 630090, Russia\\
% ...
$^{29}$ INFN Sezione di Ferrara, (A)INFN Sezione di Ferrara, I-44122, Ferrara, Italy; (B)University of Ferrara,  I-44122, Ferrara, Italy\\
% ...
$^{34}$ Johannes Gutenberg University of Mainz, Johann-Joachim-Becher-Weg 45, D-55099 Mainz, Germany\\
$^{35}$ Joint Institute for Nuclear Research, 141980 Dubna, Moscow region, Russia\\
% ...
$^{38}$ Liaoning Normal University, Dalian 116029, People's Republic of China\\
% ...
$^{55}$ Southeast University, Nanjing 211100, People's Republic of China\\
$^{56}$ State Key Laboratory of Particle Detection and Electronics, Beijing 100049, Hefei 230026, People's Republic of China\\
% ...
$^{69}$ University of Science and Technology of China, Hefei 230026, People's Republic of China\\
% ...
$^{72}$ University of Turin and INFN, (A)University of Turin, I-10125, Turin, Italy; (B)University of Eastern Piedmont, I-15121, Alessandria, Italy; (C)INFN, I-10125, Turin, Italy\\
$^{73}$ Uppsala University, Box 516, SE-75120 Uppsala, Sweden\\
% ...
$^{b}$ Also at the Novosibirsk State University, Novosibirsk, 630090, Russia\\
% ...

}

	\end{center}
}

\begin{abstract}

A precision measurement of the matrix elements for $\eta\ra\pip\pim\pio$ and $\eta\to\pio\pio\pio$
decays is performed using a sample of $(10087\pm44)\times10^6$ $\jpsi$ decays collected with the
BESIII detector. The decay $J/\psi \to \gamma \eta$ is used to select clean samples of 631,686
$\eta\ra\pip\pim\pio$ decays and 272,322 $\eta\to\pio\pio\pio$ decays.
The matrix elements for both channels are in reasonable agreement with previous
measurements. The non-zero $gX^2Y$ term for the decay mode $\eta\ra\pip\pim\pio$ is confirmed, as
reported by the KLOE Collaboration, while the other higher-order terms are found to be insignificant. 
Dalitz plot asymmetries in the $\eta\ra\pip\pim\pio$ decay are also explored and are found to be
consistent with charge conjugation invariance. In addition, a cusp effect is investigated in the
$\eta\ra\pio\pio\pio$ decay, and no obvious structure around the $\pi^+\pi^-$ mass threshold is
observed.

\end{abstract}

\maketitle

\section{Introduction}\label{sec:introduction}

The decay of the $\eta$ meson into 3$\pi$ violates isospin symmetry and
is related to the difference of light-quark masses, $m_u-m_d$, where $m_u$
and $m_d$ are the masses of valence quark $u$ and $d$, respectively.
Therefore, the decay $\eta\rightarrow3\pi$ offers a unique way to determine
the quark mass ratio $Q^2\equiv (m_s^2-{\hat m}^2)/(m_d^2-m_u^2)$, where
${\hat m} = \frac{1}{2}(m_d + m_u)$. This has stimulated both theoretical
and experimental interest~\cite{Fang:2021hyq}. Extensive theoretical studies
have been performed within the framework of combined chiral perturbation theory
(ChPT) and dispersion theory~\cite{Gasser:1984pr,Bijnens:2007pr,Schneider:2010hs,Kampf:2011wr,Guo:2015zqa,Colangelo:2018jxw}.
Experimentally, the most recent results are from the WASA-at-COSY~\cite{WASA-at-COSY:2008rsh, WASA-at-COSY:2014wpf},
KLOE/KLOE-2~\cite{KLOE:2010ytm, KLOE-2:2016zfv}, \mbox{BESIII}~\cite{BESIII:2015fid}
~(with 1.3 billion $J/\psi$ events collected in 2009 and 2012),
and A2~\cite{A2:2018pjo} Collaborations. Taking experimental results as inputs,
two dedicated analyses by independent groups of theorists reported
$Q= 22.0 \pm 0.7$~\cite{Colangelo:2016jmc} and $Q = 21.6 \pm 1.1$~\cite{Guo:2016wsi}.

In addition, a sizeable cusp structure is expected to be visible in the mass
spectra of $\pio\pio$ for $\eta\ra\pio\pio\pio$ around the $\pip\pim$ mass
threshold~\cite{Gullstrom:2008sy, Bissegger:2007yq}, corresponding to the virtual
transition $\pi^0\pi^0\rightarrow\pi^+\pi^-\rightarrow\pi^0\pi^0$~\cite{Colangelo:2018jxw}.
An analogous cusp was first predicted~\cite{Budini:1961bac} and then observed
in the $K^+ \rightarrow \pi^+ \pi^0 \pi^0$ by the NA48/2 Collaboration~\cite{NA482:2005wht}.
Furthermore, the BESIII collaboration observed the evidence of a cusp structure
with a statistical significance of around 3.5$\sigma$ in $\eta^\prime\rightarrow\eta
\pi^0 \pi^0$~\cite{BESIII:2022tas}, which is consistent with the cusp effect predicted
by the non-relativistic effective field theory (NREFT)~\cite{Kubis:2009sb}.
A series of experimental efforts have been spent to investigate the Dalitz plot of
$\eta\rightarrow\pi^0\pi^0\pi^0$ decays~\cite{WASA-at-COSY:2014wpf,BESIII:2015fid,KLOE-2:2016zfv},
but no obvious cusp effect has yet been observed. The most recent result from
the A2 Collaboration~\cite{A2:2018pjo} stated that it is necessary to introduce
the cusp effect term in order to better describe their experimental data,
but the effect included a large uncertainty of about 50\%.

With the additional $\jpsi$ data collected in 2018 and 2019, the total number
of $J/\psi$ decays accumulated with the BESIII detector has increased to
$(10087\pm44)\times10^6$~\cite{BESIII:2021cxx}, which is about 8 times larger
than the sample used in the previous BESIII analysis~\cite{BESIII:2015fid}.
This provides a unique opportunity to further investigate the decays 
$\eta\rightarrow \pi^+\pi^-\pi^0$ and $\eta\rightarrow \pi^0\pi^0\pi^0$.
In this paper, the parameterizations of the Dalitz plot amplitudes for
$\eta\rightarrow \pi^+\pi^-\pi^0$ and $\eta\rightarrow\pi^0\pi^0\pi^0$
decays follow the previous measurements and are briefly described below.

The Dalitz plot for the mode $\eta\ra\pip\pim\pio$ is generally described
by the following two variables~\cite{BESIII:2015fid}
\begin{equation}
	X=\frac{\sqrt{3}}{Q_{\eta}}(T_{\pip}-T_{\pim}),~~~~~~ Y=\frac{3T_{\pio}}{Q_\eta}-1,
\end{equation}
where $T_{\pi}$ denotes the kinetic energy of a given pion in the $\eta$ rest
frame, $Q_\eta = m_{\eta} -m_{\pip} - m_{\pim} - m_{\pio}$ is the excess energy of
the reaction, and $m_{\eta/\pi}$ are the nominal masses from the Particle Data Group
(PDG)~\cite{Workman:2022ynf}. The square of the decay amplitude can be parameterized as
\begin{eqnarray} \label{eq:chaamp}
    |A(X,Y)|^{2} \propto & ~ 1 + aY + bY^2 + cX + dX^2 + eXY  \notag \\ 
		& + fY^{3} + gX^2Y + \cdots,
\end{eqnarray}
where the parameters $a$, $b$, $c$, $d$, $\dots$ are the matrix elements
used to test theoretical predictions and fundamental symmetries.
Specifically, non-zero odd terms of $X$ ($c$ and $e$) and integrated
asymmetries~\cite{Layter:1972aq} in the Dalitz plot imply
the violation of charge conjugation symmetry.

For the decay $\eta\ra\pio\pio\pio$, the density distribution of the Dalitz plot
has threefold symmetry due to the three identical $\pio$s in the final state.
Hence, the density distribution can be parameterized using a polar variable
\begin{equation}
    Z = X^2 + Y^2 = \frac{2}{3}\sum^{3}_{i=1}\Big(\frac{3T_{i}}{Q_\eta} - 1\Big)^{2},
\end{equation}
and the square of the decay amplitdue is expanded as~\cite{Schneider:2010hs,Kampf:2011wr}
\begin{equation}\label{eq:neuamp}
    |A(X,Y)|^2 \propto 1 + 2\alpha Z + 2\beta(3X^2Y - Y^3) + 2\gamma Z^2 + \cdots,
\end{equation}
where $\alpha$, $\beta$, and $\gamma$ are the slope parameters,
$Q_\eta = m_{\eta} -3m_{\pio}$, and $T_{i}$ denotes the kinetic energy of each
$\pio$ in the $\eta$ rest frame. Due to the low energies of the final state particles,
the $\pio\pio$ rescattering in $\eta\ra\pio\pio\pio$ is expected to be dominated 
by the $S$-wave, which leads to the conventional amplitude parameterization
$|A(Z)|^2\propto 1 + 2\alpha Z$ of $\eta\ra\pio\pio\pio$, and is widely used in most
of the previous measurements~\cite{KLOE:2010ytm, BESIII:2015fid, CrystalBallatMAMI:2008pqf}.
However, the existence of the cusp effect allows non-zero contributions from
higher order terms, as was first explored by the A2 Collaboration~\cite{A2:2018pjo}.

\section{Detector and monte carlo simulation}\label{sec:detector}

The BESIII detector~\cite{BESIII:2009fln} records symmetric $e^+e^-$ collisions provided
by the BEPCII storage rings~\cite{Yu:2016cof}, which operate in the center-of-mass energy
range from 2.00 to 4.95 GeV, with a peak luminosity of $1\times 10^{33}$ cm$^{-2}$s$^{-1}$
achieved at $\sqrt{s}=3.77$ GeV. The BESIII detector has collected large data samples in
this energy region~\cite{BESIII:2020nme}. The cylindrical core of the BESIII detector
covers 93\% of the full solid angle and consists of a helium-based multilayer drift
chamber (MDC), a plastic scintillator time-of-flight system (TOF), and a CsI(TI)
electromagnetic calorimeter (EMC), which are all enclosed in a superconducting solenoidal
magnet providing a 1.0 T (0.9 T in 2012) magnetic field~\cite{Huang:2022wuo}.
The solenoid is supported by an octagonal flux-return yoke with resistive plate counter
muon identification modules interleaved with steel. The charged-particle momentum
resolution at 1 GeV/$c$ is 0.5\%, and the specific ionization energy loss $dE/dx$
resolution is 6\% for electrons from the Bhabha scattering. The EMC measures cluster
energies with a resolution of 2.5\% (5\%) at 1 GeV in the barrel (end cap) region.
The time resolution in the TOF barrel region is 68 ps, while that in the end cap
region is 110 ps. The end cap TOF system was upgraded in 2015 using multi-gap resistive
plate chamber technology, providing a time resolution of 60 ps~\cite{Litof, Cao:2020ibk}.

Simulated data samples produced with a {\sc geant4}-based~\cite{GEANT4:2002zbu} Monte Carlo (MC)
package, which includes the geometric description of the BESIII detector and the detector
response, are used to determine detection efficiencies and to estimate backgrounds.
The simulation models the beam energy spread and initial state radiation in the
$e^+e^-$ annihilation with the generator {\sc kkmc}~\cite{Jadach:2000ir,Jadach:1999vf}. 
A simulated sample of 10 billion $J/\psi$  inclusive decays (inclusive MC sample)
including both the production of the $J/\psi$ resonance and the continuum processes
is produced with {\sc kkmc} to identify background contributions. All particle decays are
modelled with {\sc evtgen}~\cite{Lange:2001uf,Ping:2008zz} using branching fractions 
either taken from the PDG~\cite{Workman:2022ynf}, when available,
or otherwise estimated with {\sc lundcharm}~\cite{Chen:2000tv,Yang:2014vra}.
Final state radiation from charged final state particles is incorporated using the
{\sc photos} package~\cite{Richter-Was:1992hxq}.

\section{ Analysis of the decay $\eta\to\pip\pim\pio$}\label{chadp}

To reconstruct $\jpsi\ra\gamma\eta$ with the subsequent decays $\eta\ra\pip\pim\pio$ and
$\pio\ra\gamma\gamma$, candidate events are required to have exactly two oppositely
charged tracks and at least three photon candidates. Charged tracks detected
in the MDC are required to be within a polar angle ($\theta$) range of
$|\cos\theta| < 0.93$, where $\theta$ is defined with respect to the $z$-axis,
which is the symmetry axis of the MDC. The distance of closest approach for
each track must be less than 10 cm along the positron beam direction and less
than 1 cm in the transverse plane with respect to the interaction point (IP).
Photon candidates, identified using isolated clusters in the EMC, are required to
have a deposited energy greater than 25 MeV in the barrel region ($ |\cos\theta| < 0.80$)
and 50 MeV in the end-cap region (0.86 $ < |\cos\theta| < $ 0.92). To eliminate
clusters originating from charged tracks, the angle subtended by the EMC cluster
and the position of the closest charged track at the EMC must be greater than 10 degrees
as measured from the IP.  The difference between the EMC time and the event start time
is required to be within $[0,700]$~ns to suppress electronic noise and energy
deposition unrelated to the event.

\begin{figure*}[!htbp]
    \centering
    \includegraphics[width=0.50\textwidth]{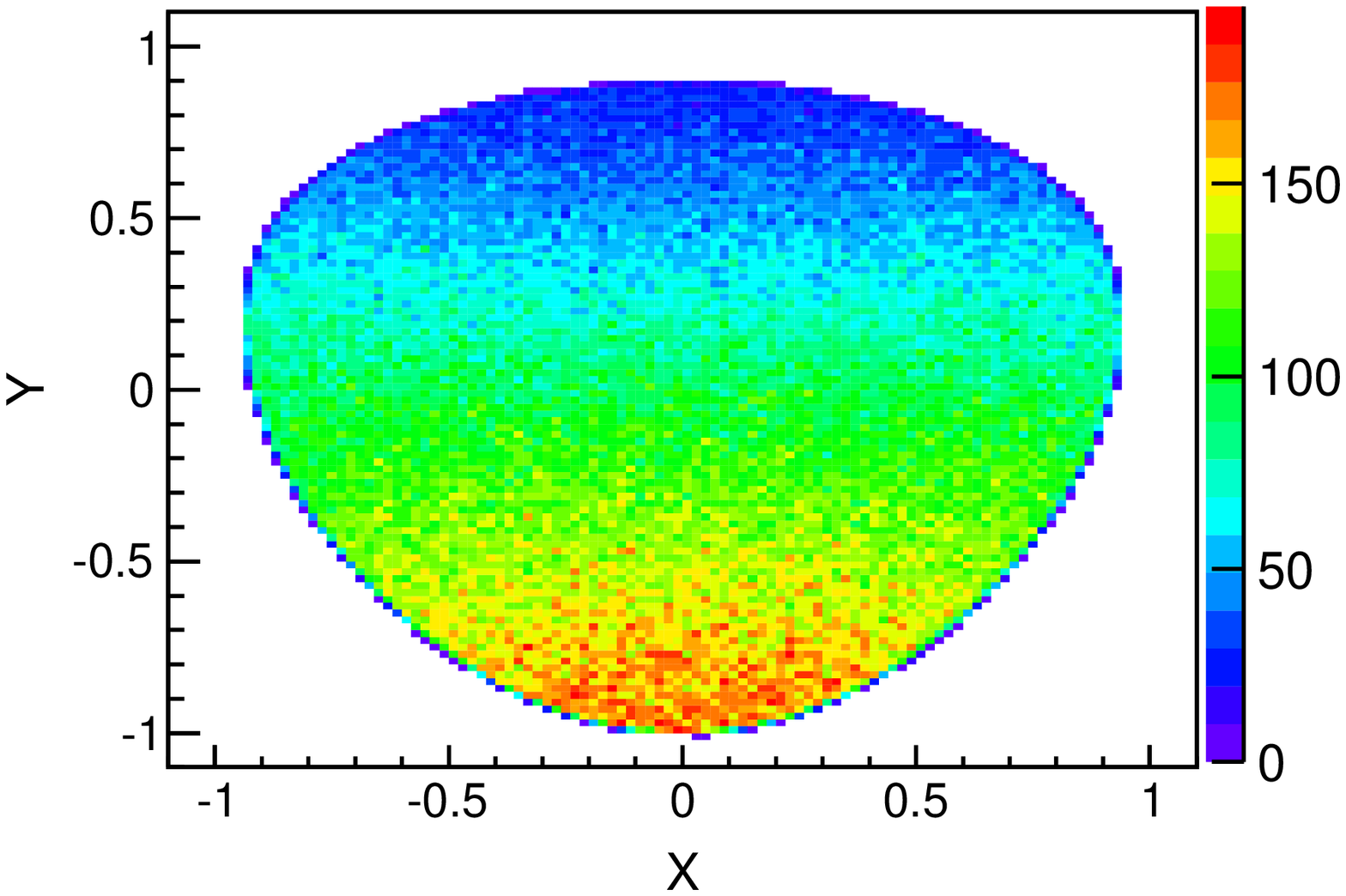}\put(-180,150){\bf (a)}
    \includegraphics[width=0.48\textwidth]{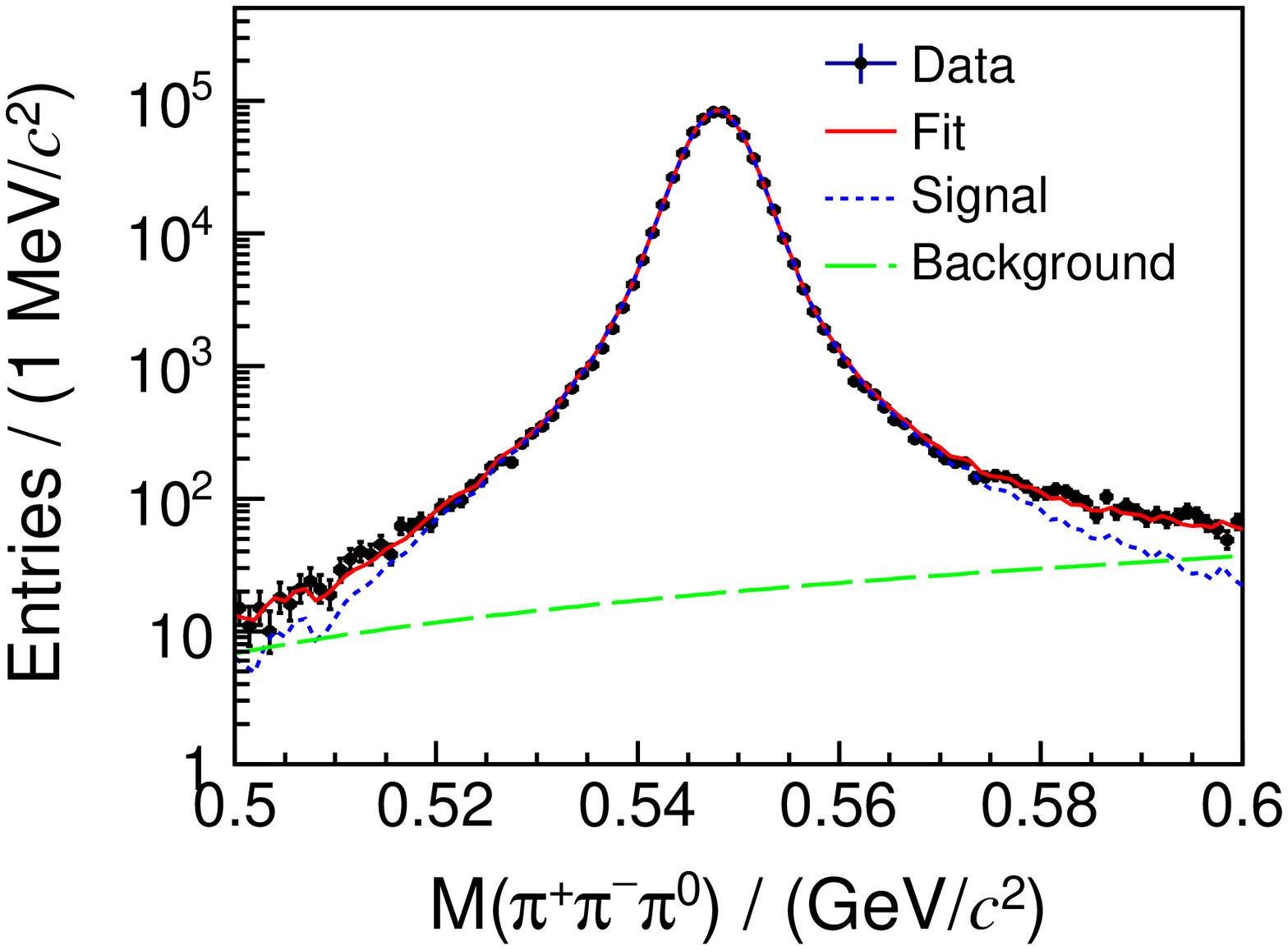}\put(-180,150){\bf (b)}
    \caption{\label{fig:evtcha} (a) Dalitz plot for $\eta\to\pip\pim\pio$ candidate events
    selected from data. (b) Invariant mass distribution of $\pip\pim\pio$ candidates without
    the $\eta$ mass constraint applied in the kinematic fit.}
\end{figure*}

Since the radiative photon from the $\jpsi$ decay is always more energetic than
the photons from the $\pio$ decay, the photon candidate with the maximum energy
in the event is taken as the radiative one. For each $\pip\pim\gamma\gamma\gamma$ combination,
a six-constraint (6C) kinematic fit is performed, and the $\chisq_{\rm 6C}$ is required
to be less than 80. The fit enforces energy-momentum conservation (4C) and constrains
the invariant masses of the other two photons and $\pip\pim\pio$ to the nominal
$\pio$ and $\eta$ masses, respectively. If there are more than three photon candidates
in an event, the combination with the smallest $\chi^{2}_{\rm 6C}$ is retained. 
Possible background events are investigated with an inclusive MC sample of 10 billion $\jpsi$ events.
To reject background events with two or four photons in the final state, we further require
that the probability of the 4C kinematic fit imposing energy-momentum conservation
for the $\jpsi\to\pip\pim\gamma\gamma\gamma$ signal hypothesis is smaller than those for
the $\jpsi\to\pip\pim\gamma\gamma$ and $\jpsi\to\pip\pim\gamma\gamma\gamma\gamma$
background hypotheses. After applying the above selection criteria,
there are no peaking backgrounds in the $\eta$ signal region and the background
contamination ratio is estimated to be 0.12\%.
A sample of 631,686 $\eta\ra\pip\pim\pio$ candidate events is selected
and the corresponding Dalitz plot of $X$ versus $Y$ is shown in Fig.~\ref{fig:evtcha}(a). 
The overall efficiency is estimated to be $(25.57\pm0.01)\%$ using a dedicated
MC simulation based on the previous BESIII measurement~\cite{BESIII:2015fid}.

To estimate the background contribution under the $\eta$ peak,
a 5C kinematic fit without the $\eta$ mass constraint is performed.
The resulting $\pip\pim\pio$ invariant mass spectrum, $M(\pip\pim\pio)$,
is shown in Fig.~\ref{fig:evtcha}(b),
where a clear $\eta$ peak is observed. An unbinned maximum likelihood fit
is then performed to the $M(\pip\pim\pio)$ distribution, where the signal is described
by the MC-simulated shape convolved with a Gaussian resolution function, and the
background contribution is described by a second-order Chebyshev polynomial function.
The background fraction is estimated to be about $0.10\%$ in the $\eta$
signal region (0.528,0.568) GeV/$c^2$, which is consistent with the estimation from
the inclusive MC sample. Therefore, the background contribution is neglected 
in the extraction of the matrix elements.

An unbinned maximum likelihood fit to data is performed to extract the free
parameters in the decay amplitude given in Eq.~(\ref{eq:chaamp}).
To account for the resolution and detection efficiency, the amplitude squared
is convolved with a function $\sigma(X,Y)$ parameterizing the resolution and
multiplied by a function $\varepsilon(X,Y)$ representing the detection efficiency.
Both functions are derived from the dedicated MC simulation based on the previous BESIII
measurement~\cite{BESIII:2015fid}. Two double Gaussian functions are used for
$\sigma(X,Y)$, 
while $\varepsilon(X,Y)$ is estimated as the average efficiency within a given
bin of $X$ and $Y$. The probability density function
${\cal P}(X,Y)$ applied to data is defined as
\begin{equation}
 {\cal P}(X,Y)=\frac{(|A(X,Y)|^2\otimes\sigma(X,Y))\cdot\varepsilon(X,Y)}
 {\int_{\rm DP}((|A(X,Y)|^2\otimes\sigma(X,Y))\cdot\varepsilon(X,Y))dXdY}.
\end{equation}
The integral over the full Dalitz plot range (DP) gives the normalization
factor in the denominator. The fit is done by minimizing the negative
log-likelihood value
\begin{equation}\label{eq:pdf}
 -\ln {\cal L} = -\sum_{i=1}^{N_{\rm event}}\ln {\cal P}(X_i,Y_i),
\end{equation}
where ${\cal P}(X_i,Y_i)$ is evaluated for the event $i$, and the sum runs
over all accepted events.

\begin{figure*}[!htbp]
	\centering
	\includegraphics[width=0.5\textwidth,height=6cm]{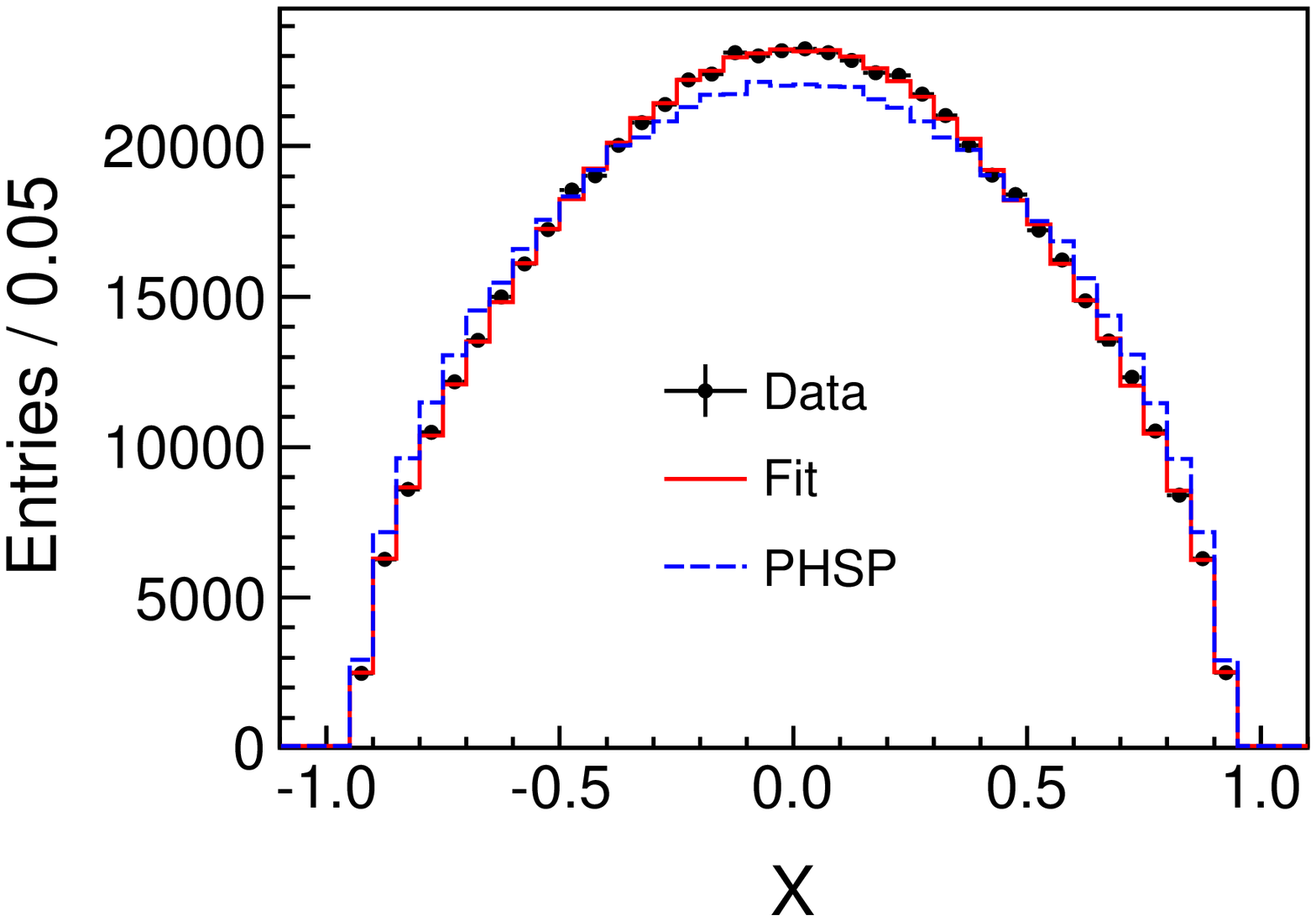}\put(-60,140){\bf (a)}
	\includegraphics[width=0.5\textwidth,height=6cm]{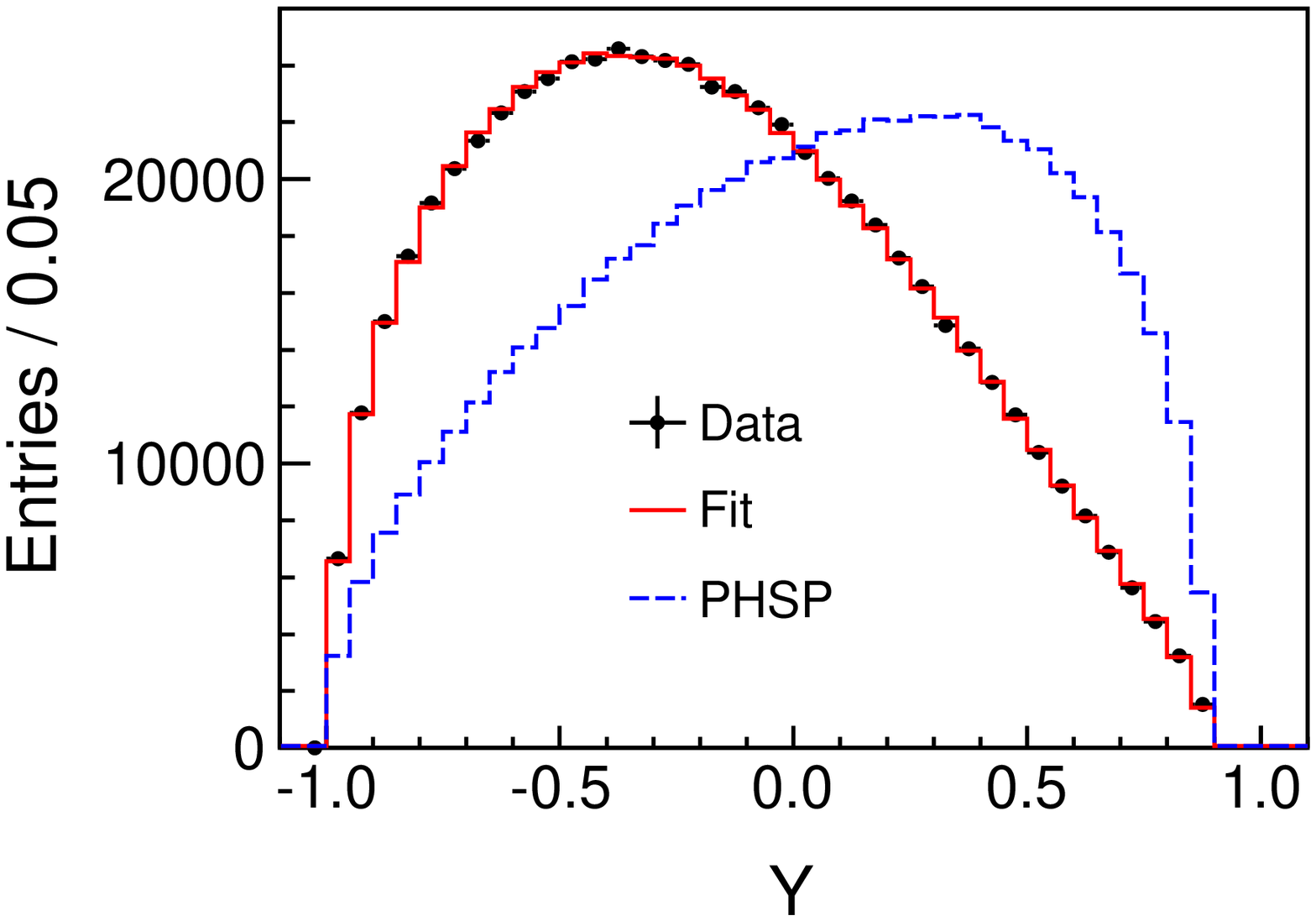}\put(-60,140){\bf (b)}
	\caption{\label{fig:etacha_dalXY_fit} Projections of the Dalitz plot on (a)~$X$ and (b)~$Y$ 
		for $\eta\ra\pip\pim\pio$ candidates, where the dots with error bars are data, the dashed
		histograms are PHSP-distributed MC events, and the solid histograms
		are the fit projections.}
\end{figure*}

Ignoring the high-order term $gX^2Y$ and imposing charge conjugation invariance 
by setting the coefficient of odd powers in $X$ ($c$ and $e$) to zero,
the fit yields the following parameters 
\begin{equation}\label{eq:chanog}
\begin{matrix} 
a & = & -1.097 &\pm& 0.005, \\
b & = & \phantom{-}0.158 &\pm& 0.006, \\
d & = & \phantom{-}0.070 &\pm& 0.006, \\
f & = & \phantom{-}0.134 &\pm& 0.010, \\
\end{matrix}
\end{equation}
and the corresponding correlation matrix
\begin{equation}
\begin{pmatrix}
   & \vline &  	a   &    b    &    d   & f     \\\hline
a  & \vline & 1.000 &  -0.262 & -0.384 & -0.751\\
b  & \vline &       & \phantom{-}1.000 &  \phantom{-}0.310 & -0.294\\
d  & \vline &       &         &  \phantom{-}1.000 &  \phantom{-}0.076 \\
f  & \vline &       &         &           &\phantom{-}1.000 \\
\end{pmatrix}
.
\end{equation}
Here the uncertainties are statistical only. The fit projections on $X$ and $Y$
are illustrated as the solid histograms in Fig.~\ref{fig:etacha_dalXY_fit}(a)
and Fig.~\ref{fig:etacha_dalXY_fit}(b). The obtained parameters are compatible
with the previous BESIII measurement~\cite{BESIII:2015fid} but the statistical
uncertainties are improved significantly. For comparison, the corresponding
distributions obtained from phase space (PHSP) distributed MC simulated events
of $\eta\ra\pip\pim\pio$ are also shown in Fig.~\ref{fig:etacha_dalXY_fit}.

An alternative fit with parameters $c$ and $e$ free is also performed to search
for the violation of charge conjugation symmetry. The fit yields 
$c = (-0.086 \pm 2.986)\times10^{-3}$ and $e = -0.001 \pm 0.007$
which are consistent with zero within statistical uncertainties, while the derived
parameters $a$, $b$, $d$ and $f$ are almost unchanged. The statistical significance
of charge-parity violation, determined by the changes of the log-likelihood value
and the number of degrees of freedom, is calculated to be $0.7\sigma$, which
indicates that there is no evidence for charge-parity violation in the decay
$\eta\ra\pip\pim\pio$.

Another fit with the cubic term $gX^2Y$ included is also performed. The fitted
parameters and corresponding correlation matrix are
\begin{equation}\label{eq:chag}
\begin{matrix} 
a & = & -1.086 &\pm& 0.006, \\
b & = & \phantom{-}0.162&\pm& 0.006, \\
d & = & \phantom{-}0.083 &\pm& 0.007, \\
f & = & \phantom{-}0.118 &\pm& 0.011, \\
g & = & -0.053 &\pm& 0.017, \\
\end{matrix}
\end{equation}
and
\begin{equation}
\begin{pmatrix}
   & \vline &  	a   &    b    &    d   & f  &     g   \\\hline
a  & \vline & 1.000 &  -0.124 & \phantom{-}0.005 & -0.785 & -0.470\\
b  & \vline &       & \phantom{-}1.000 &  \phantom{-}0.369 & -0.343 & -0.216\\
d  & \vline &       &         &  \phantom{-}1.000 &  -0.145  &  -0.595 \\
f  & \vline &       &         &           &\phantom{-}1.000  & \phantom{-}0.344\\
g  & \vline &       &         &           &           &\phantom{-}1.000 \\
\end{pmatrix}
.
\end{equation}
An evident contribution from the $X^2Y$ term is found. Based on the change
in the log-likelihood value and taking into account the change in the number
of degrees of freedom, the statistical significance of a non-zero $g$ is
calculated to be 3.0$\sigma$. The value of $g$ is in agreement with the first
and only existing experimental measurement from the KLOE-2 Collaboration,
$-0.044\pm0.009^{+0.012}_{-0.013}$~\cite{KLOE-2:2016zfv}.

The impact of the extra charge conjugation symmetry violation terms $hXY^2+lX^3$
are also investigated and the corresponding parameters are found to be
$h=-0.006\pm0.010$ and $l=-0.0009\pm0.0060$,
while the other parameters are unchanged, which means the contributions
from these terms are not significant based on the current statistics.

In addition to terms with an odd power of $X$ in Eq.~(\ref{eq:chaamp}), the unbinned
integrated asymmetries of the Dalitz plot provide a more sensitive test of charge conjugation
violation~\cite{Layter:1972aq,  Jane:1974mk, Gormley:1968zz, KLOE:2008tdy, KLOE-2:2016zfv, Gardner:2019nid, Akdag:2021efj, Akdag:2022sbn}.
The left-right asymmetry ($A_{LR}$) can be used to check charge conjugation
violation, the quadrant asymmetry ($A_Q$) is sensitive to charge conjugation violation
when isospin changes by two units, $\Delta I=2$, and the sextant asymmetry ($A_S$) is
sensitive when $\Delta I=0$. Following the same definitions as those used in the previous
measurements~\cite{KLOE:2008tdy, KLOE-2:2016zfv} and using the detection efficiency
corrected candidate events, the $A_{LR}$, $A_Q$ and $A_Q$ are calculated to be 
\begin{equation}
    \begin{array}{lc}
   A_{LR} & = \phantom{-}(0.114\pm0.131)\%,\\ 
   A_{Q}  & = (-0.035\pm0.131)\%, \\ 
   A_{S}  & = (-0.070\pm0.131)\%,
\end{array}
\end{equation}
where the uncertainties are statistical only, thereby indicating
no evidence of deviation from zero.

\section{ Analysis of the decay $\eta\ra\pio\pio\pio$}\label{sec:neudp}

To reconstruct  $\jpsi\ra\gamma\eta$ with $\eta\ra\pio\pio\pio$ and
$\pio\ra\gamma\gamma$ candidate events, at least seven photons and no charged
tracks are required. The selection criteria for the photon candidates are the
same as those described above for $\eta\ra\pip\pim\pio$, but no requirement
on the angle between photon candidates and charged tracks is imposed. To suppress electronic
noise and clusters unrelated to the event, the difference between the EMC time
and the most energetic photon is required to be within $[-500, 500]$~ns. The  
photon with maximum energy in the event is assumed to be the radiative photon
originating from the $\jpsi$ decay. For the remaining photon candidates, 
all possible $\gamma\gamma$ pairs are combined into $\pio$ candidates and subjected
separately to a one-constraint kinematic fit in which the invariant mass of the
$\gamma\gamma$ is constrained to the nominal $\pio$ mass. The $\chisq_{\pi^{0}}$
value of this kinematic fit is required to be less than 25. 
To suppress photon mis-combinations, the polar angle of the $\pio$ decay in the
$\pio$ helicity frame, defined as 
$| \cos\theta_{\rm decay} | = | E_{\gamma1} - E_{\gamma2}  | /  p_{\pio} $, 
where $E_{\gamma1}$, $E_{\gamma2}$, and $p_{\pio}$ are the photon energies and 
the $\pio$ momentum in the lab frame, respectively, is required to satisfy 
$|\cos\theta_{\rm decay}|<0.95$. Events with at least three
accepted $\pio$ candidates are retained for further analysis.

\begin{figure*}[!htbp]
    \centering
    \includegraphics[width=0.50\textwidth]{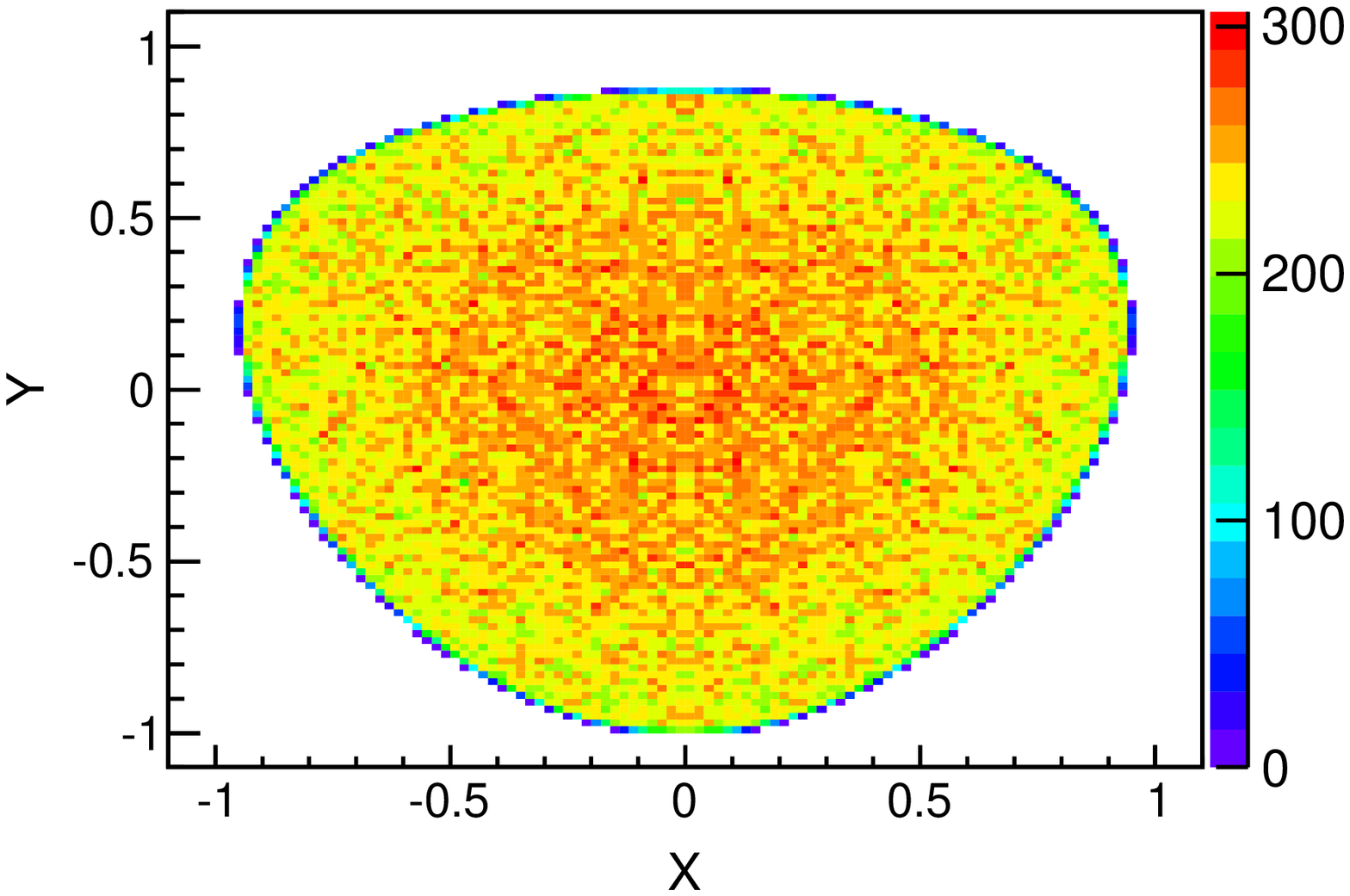}\put(-180,150){\bf (a)}
    \includegraphics[width=0.48\textwidth]{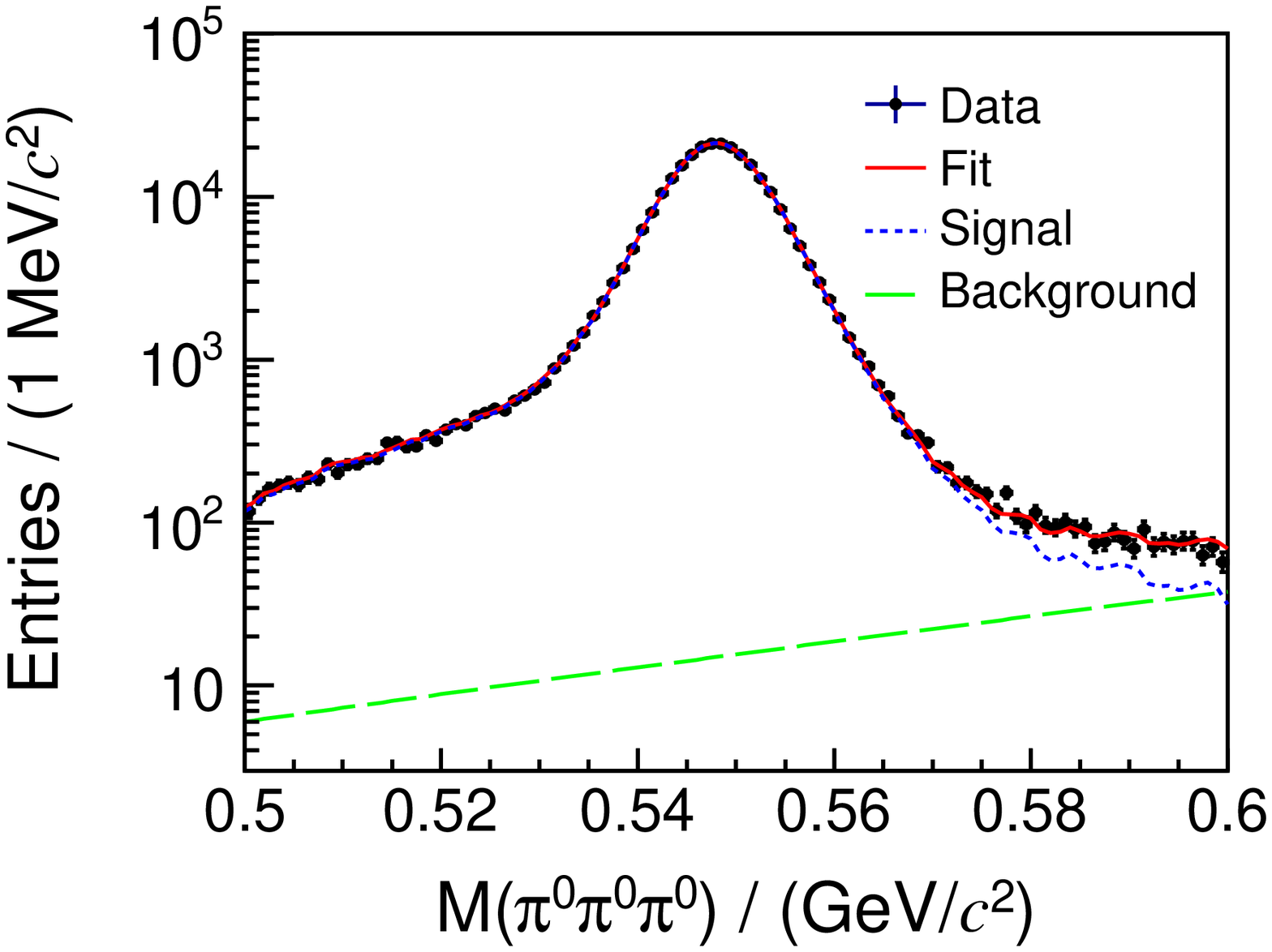}\put(-180,150){\bf (b)}
    \caption{\label{fig:evtneu} (a) Dalitz plot for $\eta\ra\pio\pio\pio$ candidate
			events selected from data (six entries per event). (b) Invariant mass spectrum
			of $\pio\pio\pio$ candidates without the $\eta$ mass constraint applied in the
			kinematic fit.
    }
\end{figure*}

\begin{figure}[!htbp]
    \centering
		\includegraphics[width=0.5\textwidth]{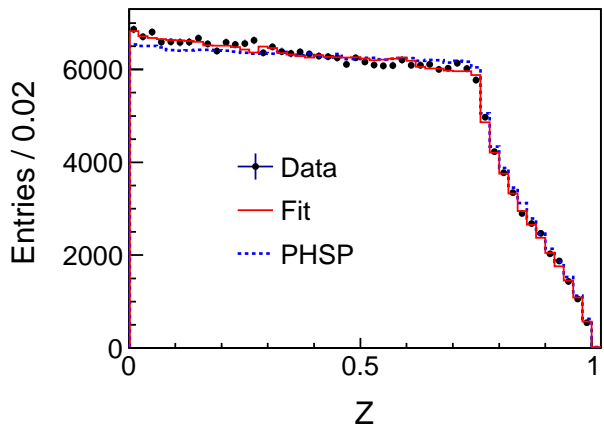}
		\caption{\label{fig:neu_dalXY_fit} Distribution of the kinematic variable $Z$ for
			$\eta\ra\pio\pio\pio$ candidates obtained from data (dots with error bars), 
			PHSP-distributed MC events (dotted histogram), and the fit projection (solid histogram). 
		}
\end{figure}

An eight-constraint (8C) kinematic fit is performed and the resulting $\chi^{2}_{\rm 8C}$ is required
to be less than 70. The 8C kinematic fit enforces energy-momentum conservation and constrains the
invariant masses of the three $\gamma\gamma$ pairs and the $\pio\pio\pio$ to the nominal $\pio$
and $\eta$ masses, respectively. If there is more than one combination in an event, 
the one with the smallest $\chi^{2}_{\rm 8C}$ is retained. 
After all selection criteria are imposed, 272,322 $\eta\ra\pio\pio\pio$ events are selected for further analysis
and the corresponding Dalitz plot of $X$ versus $Y$ is shown in Fig.~\ref{fig:evtneu}(a). 
The overall efficiency is estimated to be $(8.15\pm0.01)\%$ from the dedicated MC simulation
based on the BESIII previous measurement~\cite{BESIII:2015fid}.
As mentioned above, there are three identical $\pi^0$ in the final state, so the variables
$X$ and $Y$ can be determined in six different ways but with the same value for the variable $Z$.
Therefore, one event contributes to six entries in Fig.~\ref{fig:evtneu}(a). The distribution of
the kinematic variable $Z$ for the selected candidate events in data is shown in Fig.~\ref{fig:neu_dalXY_fit}.

To estimate the background contribution under the $\eta$ peak,
a 7C kinematic fit without the $\eta$ mass constraint is performed. 
The resulting $\pio\pio\pio$ invariant mass spectrum, $M(\pio\pio\pio)$,
is shown in Fig.~\ref{fig:evtneu}(b) with a very prominent $\eta$ peak. 
An unbinned maximum likelihood fit to the $M(\pio\pio\pio)$ distribution
is performed, where the signal is modelled by the MC simulated shape
convolved with a Gaussian resolution function, and the background contribution
is described with a second-order Chebyshev polynomial function.
The background fraction is estimated to be $0.3\%$, which
is consistent with that found with the inclusive MC sample of
10 billion $J/\psi$ events. We therefore neglect the background contribution in the
extraction of the matrix elements.

As described in Sec.~\ref{sec:introduction}, the conventional leading-order parameterization
for $\eta\ra\pio\pio\pio$ only relies on the quadratic slope parameter $\alpha$. 
Analogous to the measurement for $\eta\ra\pip\pim\pio$, an unbinned maximum likelihood
fit to data is performed on the $Z$ distribution as a function of $Z=X^2+Y^2$ to
extract the slope parameter.  After taking into account the mass resolution and
the detection efficiency, which are obtained from the MC simulation, 
the fit only considering the slope parameter up to $Z$ term in Eq.~(\ref{eq:neuamp}) gives 
\begin{equation}  \label{eq:alpha}
	\alpha = -0.0406\pm0.0035,
\end{equation}
where the uncertainty is statistical only. The fit result, as illustrated
in Fig.~\ref{fig:neu_dalXY_fit}, indicates that the decay amplitude with the
quadratic slope parameter $\alpha$ describes data well.
To directly compare with the previous BESIII measurement~\cite{BESIII:2015fid}, $-0.055\pm0.014\pm0.004$,
an alternative fit with the requirement $Z<0.7$ is also performed and yields $\alpha=-0.046\pm0.005$,
which is consistent with the previous measurement within uncertainties.

To test for higher-order contributions, an alternative fit including the terms 
$2\alpha Z$+$2\beta(3X^2Y - Y^3)$ gives  $\alpha=-0.0421\pm0.0037$ and $\beta=0.0038\pm0.0033$,
which implies that $\beta$ is consistent with zero and this extra term has little impact
on the value of $\alpha$. 
Please note that one $\eta\ra\pio\pio\pio$ event contributes to six entries in
the Dalitz plot of $X$ and $Y$, the parameter errors from this fit are multiplied by
the factor of $\sqrt6$ to reflect the actual experimental statistics.
We also perform the fits by including the terms
$2\alpha Z$+$2\gamma Z^2$ and find that $\gamma=-0.018\pm0.014$ is also consistent
with zero, but has a large impact on $\alpha$ due to a strong
correlation, $-0.964$, between $\alpha$ and $\gamma$. Therefore, with the current
sample size, it is justified to ignore
the higher-order contributions.

\begin{figure}[!htbp]
  \centering
	\includegraphics[width=0.5\textwidth]{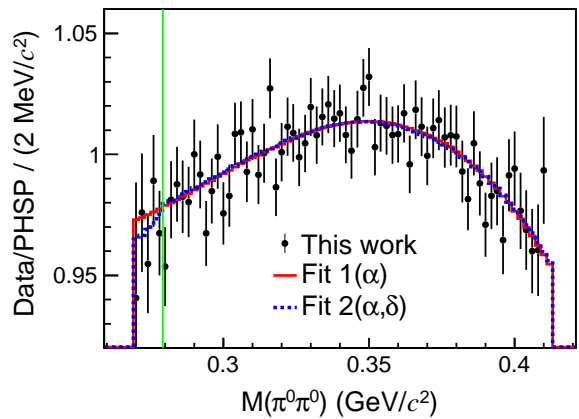}
  \caption{\label{fig:cuspfit} Ratio between the experimental $\pio\pio$ invariant mass
		distributions (the black dots with error bars) or the fit results (colored curves) and the PHSP
		simulated distributions. The Fit 1($\alpha)$ and Fit 2($\alpha, \delta$) correspond 
		to the fit projections based on the results
		in Eq.~(\ref{eq:alpha}) and Eq.~(\ref{eq:alphadelta}), respectively.
		The vertical line corresponds to the $\pi^+\pi^-$ mass threshold.
	}
\end{figure}

To search for the cusp effect near the $\pi^+\pi^-$ mass threshold, we evaluate the
ratio of the experimental $\pio\pio$ invariant-mass distribution to the PHSP-distributed 
MC sample, as depicted in Fig.~\ref{fig:cuspfit}. No obvious
structure around the $\pip\pim$ mass threshold is observed. 
Using the same approach as described in Ref.~\cite{A2:2018pjo}, which is also similar
to NREFT~\cite{Schneider:2010hs},
the cusp structure is parameterized in terms of the density function
as $\rho(s_i)=\Re\sqrt{1-s_i/4m^2_{\pi^\pm}}$, where $\Re$ is the real part, $i=1,2,3$
denotes the three $\pi^0$'s, and $s_i=(p_\eta-p_i)^2$ with $p_i^2=m_i^2$.
Here $p_\eta$ and $p_i$ are the four-momenta of the $\eta$ and the $i^{\rm th}$ $\pi^0$,
respectively, while $m_i$ is the $\pi^0$ invariant mass.
For $s\ge4m^2_{\pi^\pm}$, $\rho(s)=0$. Then the amplitude squared is given by
\begin{equation}\label{eq:neucusp}
    |A|^2 \propto 1 + 2\alpha Z + 2\delta\sum^3_{i=1}{\rho(s_i)},
\end{equation}
where the factor of $2$ in front of the cusp term $\delta$ is added to be consistent
with the $Z$ term. 
An unbinned maximum likelihood fit to data on the $Z$ and $\sum^3_{i=1}{\rho(s_i)}$ 
distributions yields the folloowing parameters
\begin{equation}  \label{eq:alphadelta}
   \alpha = -0.0397\pm 0.0036,
	 \delta = -0.018 \pm 0.022,
\end{equation}
where the uncertainties are statistical only. The correlation coefficient between
$\alpha$ and $\delta$ is $-0.304$. The fitted value for $\delta$ is consistent with
zero within the statistical uncertainty and the parameter $\alpha$ is almost unchanged,
which indicates there is no evidence for a
significant cusp effect in $\eta\ra\pio\pio\pio$ decays. Comparing the log-likelihood values and
the number of degrees of freedom of the fits with ($\alpha$, $\delta$) and $\alpha$ only,
the statistical significance for non-zero $\delta$ is determined to be 0.8$\sigma$.
The fit result is also illustrated in Fig.~\ref{fig:cuspfit}.

\section{Systematic Uncertainties}\label{sec:syst}

Various sources of systematic uncertainties on the measured Dalitz plot matrix elements
have been investigated, including tracking efficiency, $\pio$ reconstruction, kinematic fit,
efficiency binning, fit bias, and different resolution between data and MC simulation.
For $\eta\ra\pio\pio\pio$ candidates, an additional uncertainty due to the photon 
mis-combination is also considered.

Differences between the data and MC samples for the tracking efficiency of charged pions
are investigated using a control sample of $\jpsi\ra\pip\pim\pio$ events. 
A transverse momentum and cos$\theta$ dependent correction to the detection efficiency is
obtained by comparing the efficiency between the data and MC simulation, where $\theta$ is
the polar angle of the track. Similarly, a momentum dependent correction for  
$\pio$ reconstruction is investigated using $\jpsi\ra\pip\pim\pio$ decays.  
Alternative fits are performed by considering the efficiency corrections for charged
pions or $\pio$, and the changes of the matrix elements
with respect to the nominal results are taken as the systematic uncertainties.

Uncertainties associated with the kinematic fits mainly come from the inconsistency of the
track helix parameters and photon resolution between data and MC simulation.
For $\eta\to\pip\pim\pio$, the helix parameters and error matrix for the charged tracks 
in the MC samples are
corrected to eliminate the inconsistency between data and MC simulation.
While for $\eta\to\pio\pio\pio$ candidates, the raw photon energy from the EMC 
and the uncertainty of the photon energy in the 
kinematic fit are adjusted to reduce the difference between data and MC simulation.
Alternative fits are performed with the corrected MC samples, 
and the corresponding differences on the matrix elements are taken as the systematic
uncertainties.

Different binning schemes of the efficiency are exploited to estimate the associated
uncertainties. The changes of the matrix elements with respect to the default fit results are
assigned as the systematic uncertainties.

To estimate the systematic uncertainties due to the fitting procedure,
input-output checks are performed with the dedicated MC samples. The deviations
of the pull distributions from the standard normal distributions are 
taken as the systematic uncertainties.

The possible mis-combination of photons in the $\eta\to\pio\pio\pio$ sample has been studied by
matching the generated photon pairs to the selected $\pio$ candidates. The fraction
of events with mis-combined photons is determined to be $5.3\%$.
Alternative fits to the dedicated $\eta\to\pio\pio\pio$ MC sample are performed with
and without removing the mis-combined events individually. The difference between
the two fitting results is considered as a source of systematic uncertainty.

To evaluate the impact from the different resolutions of Dalitz plot variables
between data and MC samples, alternative fits are performed by varying the resolution
$\pm10\%$. 
The changes of the matrix elements with respect to the default fit results are
assigned as the systematic uncertainties.
The impact from the background contribution is also checked and the change of the
results is found to be negligible.

All of the above contributions are summarized in Table~\ref{tab:syserr}, where the total
systematic uncertainties are given by the quadratic sum of the individual uncertainties,
assuming all the sources are independent.

For the measurement of the asymmetries in the $\eta\rightarrow\pi^+\pi^-\pi^0$ Dalitz plot,
the systematic uncertainties are mainly from the kinematic fit and the difference of
tracking efficiency and $\pi^0$ reconstruction
efficiency between data and MC simulation, as summarized in Table~\ref{tab:syserr}.
To estimate the uncertainties associated with the kinematic fit, the track helix parameters
and error matrix of MC simulation are first corrected to reduce the difference between data
and MC simulation. The efficiencies are then re-estimated with the corrected MC samples,
and the differences on the asymmetries are taken as the systematic uncertainties.
With the same method described above, the differences between data and MC samples
for the tracking efficiency of charged pions and $\pi^0$ reconstruction are considered in the estimation
of the efficiencies, and the differences in the asymmetries are taken as the systematic uncertainties. 

\begin{table*}[htpb]
\centering
 \caption{\label{tab:syserr} Relative systematic uncertainties of the matrix elements for $\eta\ra\pip\pim\pio$ and $\eta\ra\pio\pio\pio$ decays, and the Dalitz plot asymmetries for $\eta\ra\pip\pim\pio$ decays (all values are given in $\%$).}
 \begin{tabular}{c|cccc|ccccc|c|ccc}\hline\hline
 \multirow{2}{*}{Source} & \multicolumn{4}{|c|}{Results in Eq.~(\ref{eq:chanog})} & \multicolumn{5}{c|}{Results in Eq.~(\ref{eq:chag})} & \multirow{2}{*}{$\alpha$} & \multirow{2}{*}{$A_{LR}$} & \multirow{2}{*}{$A_Q$} & \multirow{2}{*}{$A_S$}  \\\cline{2-10}
                	     & $a$ & $b$ & $d$ & $f$ & $a$ & $b$ & $d$ & $f$ & $g$ &  &  &  &  \\ \hline
Tracking efficiency    & 0.01 & 0.02 & 0.12 & 0.02 & 0.01 & 0.04 & 0.01 & 0.02 & 0.02 &  -    & 0.07 & 0.08 & 0.07 \\
$\pio$ efficiency      & 0.03 & 0.18 & 0.08 & 0.07 & 0.02 & 0.13 & 0.01 & 0.04 & 0.16 & 0.23  & 0.01 & 0.18 & 0.13 \\
Kinematic fit          & 0.07 & 2.06 & 0.14 & 1.96 & 0.06 & 1.98 & 0.18 & 2.29 & 0.57 & 0.01  & 0.73 & 32.3 & 12.7 \\
Efficiency binning& 0.04 & 0.29 & 0.42 & 0.11 & 0.01 & 0.33 & 0.76 & 0.53 & 3.04 & 0.15  & -  & - & - \\
Fit bias  	           & 0.02 & 0.65 & 1.06 & 0.90 & 0.02 & 0.13 & 0.17 & 0.17 & 3.50 & 1.80  & -  & - & - \\
Resolution             & 0.01 & 0.08 & 0.09 & 0.02 & 0.01 & 0.03 & 0.01 & 0.01 & 0.09 & 0.02  & -  & - & - \\
Photon mis-combination  &  -   &  -   &  -   &   -  &   -  &   -  &   -  &   -  &   -  & 0.75  & -  & - & - \\\hline
Total                  & 0.09 & 2.19 & 1.16 & 2.16 & 0.07 & 2.02 & 0.80 & 2.36 & 4.67 & 1.97  & 0.74 & 32.3 & 12.8 \\\hline\hline
 \end{tabular}
\end{table*}

\section{Summary}

Using a sample of $(10087\pm44)\times10^6$ $\jpsi$ decays collected by the BESIII detector,
very clean samples of 631,686 $\eta\ra\pip\pim\pio$ events and 272,322 $\eta\ra\pio\pio\pio$
events are selected from $\jpsi\ra\gamma\eta$ radiative decays. The matrix elements for the
decays $\eta\ra\pip\pim\pio$ and $\eta\ra\pio\pio\pio$ have been determined
precisely, and supersede the previous BESIII measurement~\cite{BESIII:2015fid}, which is
based on a subsample of the present data.

\begin{figure*}[!htbp]
    \centering
		\includegraphics[width=0.48\textwidth,height=6cm]{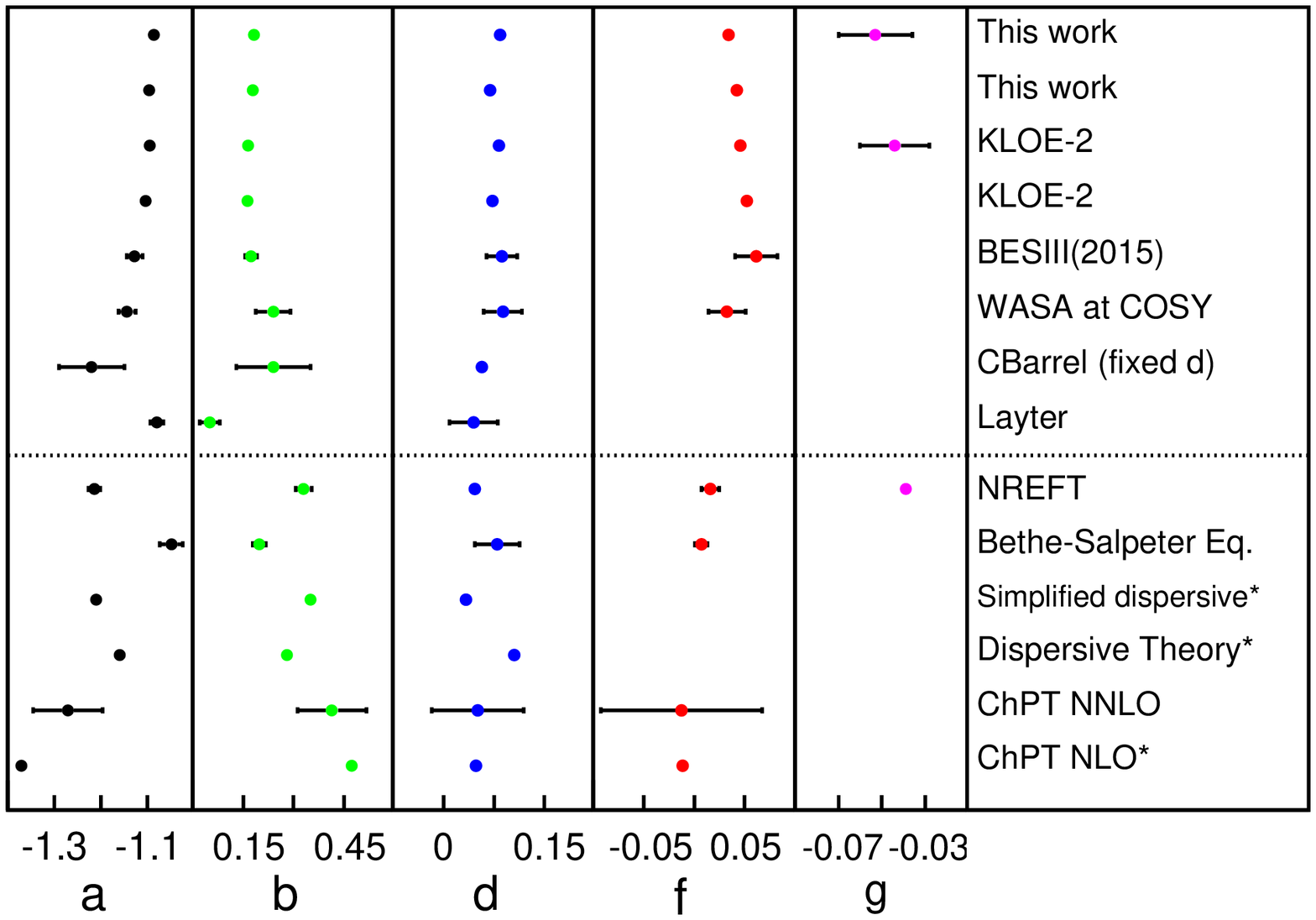}\put(-25,150){\bf (a)}
					\put(-44,138){\tiny{~\cite{KLOE-2:2016zfv}}}
					\put(-44,128.5){\tiny{~\cite{KLOE-2:2016zfv}}}
					\put(-31,119){\tiny{~\cite{BESIII:2015fid}}}
					\put(-22,110){\tiny{~\cite{WASA-at-COSY:2014wpf}}}
					\put(-21,100){\tiny{~\cite{CrystalBarrel:1998tio}}}
					\put(-49,91){\tiny{~\cite{Layter:1973ti}}}
					\put(-46,79){\tiny{~\cite{Schneider:2010hs}}}
					\put(-16,69){\tiny{~\cite{Borasoy:2005du}}}
					\put(-16,60){\tiny{~\cite{Bijnens:2002qy}}}
					\put(-16,51){\tiny{~\cite{Kambor:1995yc}}}
					\put(-31,42){\tiny{~\cite{Bijnens:2007pr}}}
					\put(-33,33){\tiny{~\cite{Bijnens:2007pr}}}
		\includegraphics[width=0.48\textwidth,height=6cm]{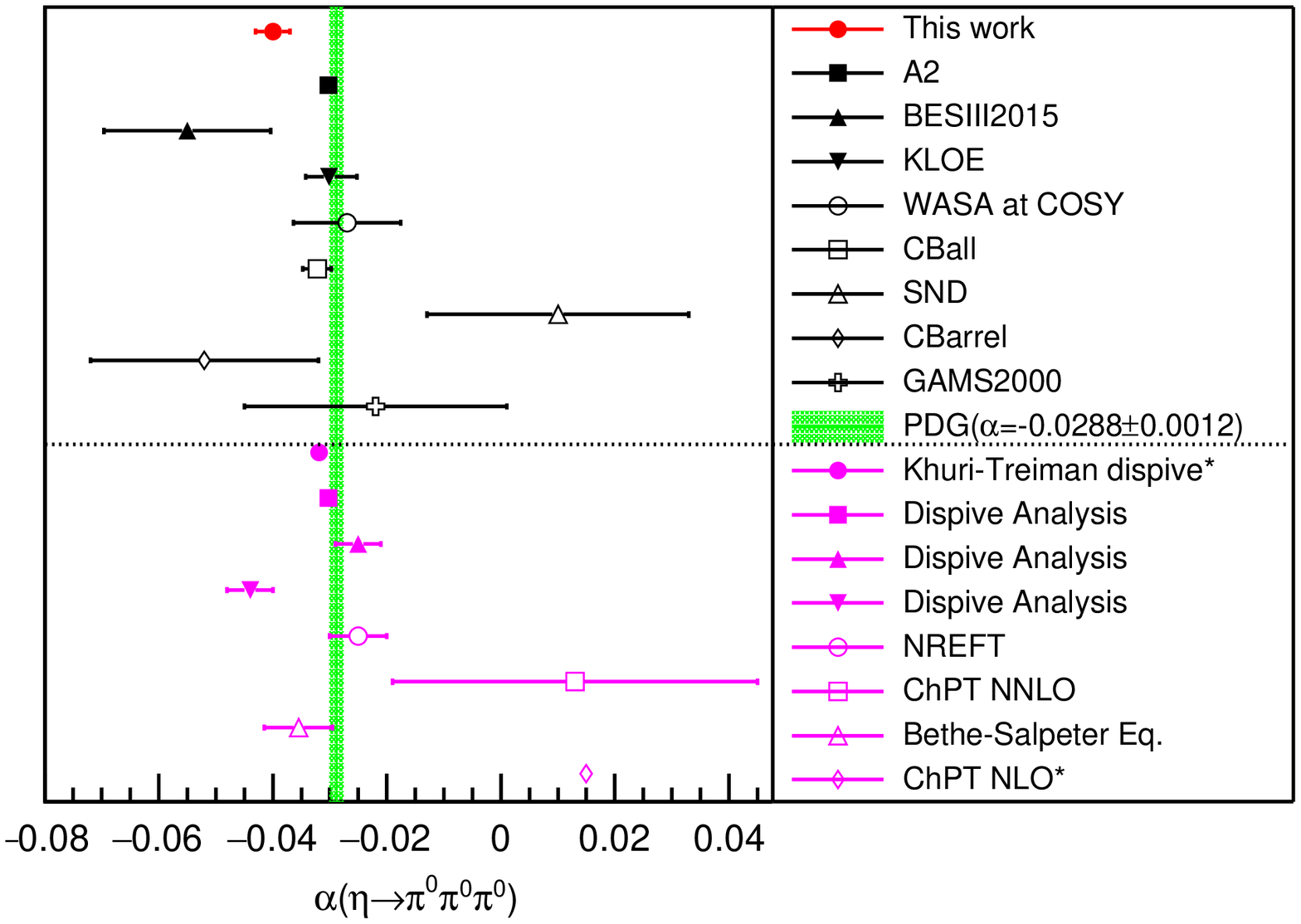}\put(-28,150){\bf (b)}
					\put(-70,149){\tiny{~\cite{A2:2018pjo}}}
					\put(-48,142){\tiny{~\cite{BESIII:2015fid}}}
					\put(-62,134){\tiny{~\cite{KLOE:2010ytm}}}
					\put(-35,127){\tiny{~\cite{WASA-at-COSY:2008rsh}}}
					\put(-64,119){\tiny{~\cite{CrystalBallatMAMI:2008pqf}}}
					\put(-65,112){\tiny{~\cite{Achasov:2001xi}}} 
					\put(-58,104){\tiny{~\cite{CrystalBarrel:1998okb}}}
					\put(-45,97){\tiny{~\cite{Serpukhov-Brussels-AnnecyLAPP:1984udf}}} 
					\put(-16,89){\tiny{~\cite{Workman:2022ynf}}}
					\put(-20,81){\tiny{~\cite{Albaladejo:2017hhj}}}
					\put(-36,73){\tiny{~\cite{Colangelo:2016jmc}}} 
					\put(-36,66){\tiny{~\cite{Guo:2016wsi}}} 
					\put(-36,59){\tiny{~\cite{Kampf:2011wr}}} 
					\put(-60,51){\tiny{~\cite{Schneider:2010hs}}} 
					\put(-45,43){\tiny{~\cite{Bijnens:2007pr}}}
					\put(-30,36){\tiny{~\cite{Borasoy:2005du}}}
					\put(-47,28){\tiny{~\cite{Bijnens:2002qy}}}
    \caption{\label{fig:resultcom} Comparison of the experimental measurements and
			theoretical calculations for  (a) $\eta\to\pip\pim\pio$ and (b) $\eta\to\pio\pio\pio$.
			No uncertainties were reported for theoretical predictions marked by $*$.}
\end{figure*}

Including only the charge conjugation invariant terms from Eq.~(\ref{eq:chaamp}),
the Dalitz plot matrix elements for $\eta\ra\pip\pim\pio$ are determined to be
\begin{equation}
\begin{matrix} 
a & = & -1.097 &\pm& 0.005 &\pm& 0.001, \\
b & = & \phantom{-}0.158 &\pm& 0.006 &\pm& 0.003, \\
d & = & \phantom{-}0.070 &\pm& 0.006 &\pm& 0.001, \\
f & = & \phantom{-}0.134 &\pm& 0.010 &\pm& 0.003, \\
\end{matrix}
\end{equation}
where the first uncertainties are statistical and the second systematic, here
and in the following.
Including the cubic term $gX^2Y$ results in the following matrix elements
\begin{equation}
\begin{matrix} 
a & = & -1.086 &\pm& 0.006 &\pm& 0.001, \\
b & = & \phantom{-}0.162 &\pm& 0.006 &\pm& 0.003, \\
d & = & \phantom{-}0.083 &\pm& 0.007 &\pm& 0.001, \\
f & = & \phantom{-}0.118 &\pm& 0.011 &\pm& 0.003, \\
g & = & -0.053 & \pm & 0.017 & \pm & 0.003.\\
\end{matrix}
\end{equation}
Our results are consistent with recent experimental
results~\cite{WASA-at-COSY:2014wpf, KLOE-2:2016zfv, BESIII:2015fid},
and are in reasonable agreement with the theoretical calculation based on the
dispersive approach and ChPT at next-to-next-to-leading order (NNLO)
level~\cite{Bijnens:2007pr,Kambor:1995yc,Bijnens:2002qy,Schneider:2010hs},
as summarized in Fig.~\ref{fig:resultcom}(a).

Charge conjugation invariance in $\eta\rightarrow\pi^+\pi^-\pi^0$ is investigated by
performing an alternative fit with odd powers of $X$ ($c$ and $e$) left free and checking the
integrated asymmetries of the Dalitz plot distribution. The obtained values,
$A_{LR}= (0.114\pm0.131\pm0.001)\%$, $A_{Q}=(-0.035\pm0.131\pm0.011)\%$ and
$A_{S}= (-0.070\pm0.131\pm0.009)\%$, are consistent
with zero and indicate no evidence for significant charge symmetry breaking.
Comparison of the experimental asymmetries results are summarized in Table~\ref{tab:ComAsy}.
In addition, the higher order contributions $hXY^2+lX^3$ are also
checked, but no significant contribution is found.

The slope parameter for $\eta\ra\pio\pio\pio$ is determined to be 
$\alpha= -0.0406\pm0.0035\pm0.0008$, which is consistent with the 
A2 measurement~\cite{A2:2018pjo}, $-0.0302\pm0.0008(\rm stat.)$, within $2.8\sigma$.
The A2's result is the most precise measurement and from the fit 
with $\alpha$ as the only free paramter.
Comparison between the theoretical calculations and experimental
measurements of the slope parameters are shown in Fig.~\ref{fig:resultcom}(b).
No significant higher-order contribution is found at the current level of precision. 
In addition, the cusp effect in $\eta\ra\pio\pio\pio$ is also investigated,
but no obvious contribution is found.

\begin{table*}[!htbp]
	\centering
	\caption{\label{tab:ComAsy} Experimental asymmetries in the Dalitz plot of $\eta\ra\pip\pim\pio$ decays.}
	\begin{tabular}{c|ccc}\hline\hline
		Experiment	    & $A_{LR} (\%)$ & $A_Q (\%)$ & $A_S (\%)$ \\\hline
		This work                     & $0.114\pm0.131\pm0.001$ & $-0.035\pm0.131\pm0.011$ & $-0.070\pm0.131\pm0.009$    \\
		KLOE-2~\cite{KLOE-2:2016zfv}  & $-0.050\pm0.045^{+0.050}_{-0.110}$ & $0.018\pm0.045^{+0.048}_{-0.023}$ & $0.004\pm0.045^{+0.031}_{-0.035}$ \\
		Jane~\cite{Jane:1974mk}       & $0.28\pm0.26$ & $-0.30\pm0.25$ & $0.20\pm0.25$    \\
		Layter~\cite{Layter:1972aq}   & $-0.05\pm0.22$ & $-0.07\pm0.22$ & $0.10\pm0.22$    \\
		Gormley~\cite{Gormley:1968zz} & $1.5\pm0.5$ & - & $0.5\pm0.5$   \\\hline\hline
	\end{tabular}
\end{table*}

\begin{acknowledgments}

The BESIII Collaboration thanks the staff of BEPCII and the IHEP computing center for their strong support. This work is supported in part by National Key R\&D Program of China under Contracts Nos. 2020YFA0406300, 2020YFA0406400; National Natural Science Foundation of China (NSFC) under Contracts Nos. 12005195, 12225509, 11635010, 11735014, 11835012, 11935015, 11935016, 11935018, 11961141012, 12022510, 12025502, 12035009, 12035013, 12061131003, 12192260, 12192261, 12192262, 12192263, 12192264, 12192265; the Chinese Academy of Sciences (CAS) Large-Scale Scientific Facility Program; the CAS Center for Excellence in Particle Physics (CCEPP); Joint Large-Scale Scientific Facility Funds of the NSFC and CAS under Contract No. U1832207; CAS Key Research Program of Frontier Sciences under Contracts Nos. QYZDJ-SSW-SLH003, QYZDJ-SSW-SLH040; 100 Talents Program of CAS; The Institute of Nuclear and Particle Physics (INPAC) and Shanghai Key Laboratory for Particle Physics and Cosmology; ERC under Contract No. 758462; European Union's Horizon 2020 research and innovation programme under Marie Sklodowska-Curie grant agreement under Contract No. 894790; German Research Foundation DFG under Contracts Nos. 443159800, 455635585, Collaborative Research Center CRC 1044, FOR5327, GRK 2149; Istituto Nazionale di Fisica Nucleare, Italy; Ministry of Development of Turkey under Contract No. DPT2006K-120470; National Research Foundation of Korea under Contract No. NRF-2022R1A2C1092335; National Science and Technology fund; National Science Research and Innovation Fund (NSRF) via the Program Management Unit for Human Resources \& Institutional Development, Research and Innovation under Contract No. B16F640076; Polish National Science Centre under Contract No. 2019/35/O/ST2/02907; Suranaree University of Technology (SUT), Thailand Science Research and Innovation (TSRI), and National Science Research and Innovation Fund (NSRF) under Contract No. 160355; The Royal Society, UK under Contract No. DH160214; The Swedish Research Council; U. S. Department of Energy under Contract No. DE-FG02-05ER41374.
%

\end{acknowledgments}

\appendix*
\section{Acceptance corrected data}

In order to directly taken as input for theoretical calculations, the tables
containing Dalitz plot acceptance corrected data and the corresponding 
statistical uncertainties are provided in the HEPData database.
To ensure that the results obtained from the acceptanced data are consistent with those from
the unbinned likelihood fit in this analysis, we also performed alternative fits with the
least square method.

The Dalitz plot matrix elements for $\eta\ra\pip\pim\pio$ and $\eta\ra\pio\pio\pio$
are extracted from a fit to the acceptance corrected data by minimizing 
\begin{equation}\label{eq:chisqfit}
 \chi^2 = \sum_{i=1}^{X_{bin}}\sum_{j=1}^{Y_{bin}}\left(\frac{N_{ij}^{cor}-N_{the}^{ij}}{\sigma_{ij}}\right)^2
\end{equation}
where the sum runs over all the bins and $N_{the}^{ij}=\int\int|A(X,Y)|^2dX_idY_j$.
The acceptance corrected signal content in each bin of the Dalitz plot
$N_{ij}^{cor}$ is obtained by dividing event number, $N_{ij}$,
by the corresponding acceptance, $\varepsilon_{ij}$. The acceptance is obtained from
the signal MC in that bin.
The statistical uncertainty $\sigma_{ij}$ includes contributions from the experimental
data and the efficiency.

Ignoring the high-order term $gX^2Y$ and imposing charge conjugation invariance, 
the Dalitz plot matrix elements for
$\eta\ra\pip\pim\pio$ extracted from the acceptance corrected data are
\begin{equation}
\begin{matrix} 
a & = & -1.096 &\pm& 0.006, \\
b & = & \phantom{-}0.157 &\pm& 0.006, \\
d & = & \phantom{-}0.069 &\pm& 0.006, \\
f & = & \phantom{-}0.131 &\pm& 0.012, \\
\end{matrix}
\end{equation}
and the corresponding correlation matrix
\begin{equation}
\begin{pmatrix}
   & \vline &  	a   &    b    &    d   & f     \\\hline
a  & \vline & 1.000 &  -0.240 & -0.405 & -0.783\\
b  & \vline &       & \phantom{-}1.000 &  \phantom{-}0.323 & -0.292\\
d  & \vline &       &         &  \phantom{-}1.000 &  \phantom{-}0.107 \\
f  & \vline &       &         &           &\phantom{-}1.000 \\
\end{pmatrix}
.
\end{equation}
The uncertainties are statistical only, here and in the following. 
The results are consistent within statistical uncertainties with the values obstained
using the unbinned maximum likelihood fit. 
Including the cubic term $gX^2Y$ results in the following matrix elements
\begin{equation}
\begin{matrix} 
a & = & -1.085 &\pm& 0.007, \\
b & = & \phantom{-}0.162&\pm& 0.007, \\
d & = & \phantom{-}0.083 &\pm& 0.008, \\
f & = & \phantom{-}0.114 &\pm& 0.013, \\
g & = & -0.056 &\pm& 0.019, \\
\end{matrix}
\end{equation}
and corresponding correlation matrix
\begin{equation}
\begin{pmatrix}
   & \vline &  	a   &    b    &    d   & f  &     g   \\\hline
a  & \vline & 1.000 &  -0.068 & \phantom{-}0.046 & -0.825 & -0.518\\
b  & \vline &       & \phantom{-}1.000 &  \phantom{-}0.401 & -0.362 & -0.251\\
d  & \vline &       &         &  \phantom{-}1.000 &  -0.181  &  -0.615 \\
f  & \vline &       &         &           &\phantom{-}1.000  & \phantom{-}0.419\\
g  & \vline &       &         &           &           &\phantom{-}1.000 \\
\end{pmatrix}
.
\end{equation}
While the slope parameter for $\eta\ra\pio\pio\pio$ is determined to be 
\begin{equation} 
	\alpha = -0.0409\pm0.0038.
\end{equation}
Please note that one $\eta\ra\pio\pio\pio$ event contributes to six entries in
the Dalitz plot of $X$ and $Y$, the parameter error from fitting to the
$\eta\ra\pio\pio\pio$ acceptance corrected data must be multiplied by
the factor of $\sqrt6$ to reflect the actual experimental statistics.


\end{document}